\documentclass[aps,prl,twocolumn,reprint,a4paper,nofootinbib,amsmath,amssymb,floats,floatfix]{revtex4-1}

\usepackage{blindtext}
\usepackage{graphicx}
\usepackage{enumitem}
\usepackage{amsmath}
\usepackage{booktabs}
\usepackage{array}
\usepackage{color}
\usepackage{amssymb}
\usepackage{mathrsfs}
\usepackage[T1]{fontenc}
\usepackage[latin1]{inputenc}
\usepackage[table]{xcolor}  
\usepackage{verbatim}
\usepackage{rotating}
\usepackage{afterpage}
\usepackage{dcolumn}
\usepackage{bm}
\usepackage[labelfont=sf,font+=small,format=hang,justification=RaggedRight]{subcaption}
\renewcommand{\d}{\mathrm{d}}
\newcommand{\e}[1]{\mathrm{e}^{{#1}}}

\newcommand{\vect}[1]{\bm{\mathrm{{#1}}}}

\newcommand{\ipleft}{\langle\kern-0.2em\langle}
\newcommand{\ipright}{\rangle\kern-0.2em\rangle}

\newcommand{\fNL}{f_{\mathrm{NL}}}
\newcommand{\fNLlocal}{\fNL^\text{local}}
\newcommand{\fNLequi}{\fNL^\text{equi}}

\newcommand{\hatfNL}{\hat{f}_{\mathrm{NL}}}
\newcommand{\hatfNLlocal}{\hatfNL^\text{local}}
\newcommand{\hatfNLequi}{\hatfNL^\text{equi}}
\newcommand{\hatfNLortho}{\hatfNL^\text{ortho}}

\newcommand{\gNL}{g_{\mathrm{NL}}}

\newcommand{\Mp}{M_{\mathrm{P}}}
\newcommand{\Mpc}{\text{Mpc}}

\newcommand{\vectkL}{\vect{k}_{\mathrm{L}}}
\newcommand{\hatvectkL}{\hat{\vect{k}}_{\mathrm{L}}}
\newcommand{\kL}{k_{\mathrm{L}}}

\newcommand{\dimlessP}{\mathcal{P}}

\newcommand{\isomode}{\sigma}

\newcommand{\xls}{x_{\text{ls}}}

\newcommand{\semibold}[1]{{\fontseries{b}\selectfont{#1}}}
\newcommand{\para}[1]{\par\vspace{2mm}\noindent\semibold{{#1.}---}\ignorespaces}

\newcolumntype{s}{>{$\displaystyle}l<{$}}
\newcolumntype{t}{>{$\displaystyle}c<{$}}
\newcolumntype{u}{>{$\displaystyle}r<{$}}
\newcolumntype{v}{>{$\displaystyle}m{4cm}<{$}}

\newcolumntype{d}{D{!}{\;\pm\;}{-1}}

\begin{document}
\title{Implications of the CMB power asymmetry for the early universe}
\author{Christian T. \surname{Byrnes}}
\email{C.Byrnes@sussex.ac.uk}
\author{Donough \surname{Regan}}
\email{D.Regan@sussex.ac.uk}
\author{David \surname{Seery}}
\email{D.Seery@sussex.ac.uk}
\author{Ewan R. M. \surname{Tarrant}}
\affiliation{\small
Astronomy Centre, University of Sussex,
Falmer, Brighton BN1 9QH, UK}

\begin{abstract}
Observations of the microwave background fluctuations
suggest
a scale-dependent amplitude asymmetry of roughly $2.5\sigma$ significance.
Inflationary explanations for this `anomaly' require non-Gaussian
fluctuations which couple observable modes to those on much larger scales.
In this Letter we describe an analysis of such scenarios
which significantly extends previous treatments.
We identify the non-Gaussian `response function' which characterizes the
asymmetry,
and show that it is non-trivial to construct a model which yields
a sufficient amplitude:
many
independent fine tunings are required, often making such models
appear less likely than the anomaly they seek to explain.
We present an explicit model satisfying observational constraints
and determine for the first time how large its bispectrum would appear to
a Planck-like experiment.
Although this model is merely illustrative, we expect it is a good proxy
for the bispectrum in a sizeable class of models which generate a
scale-dependent response using a large $\eta$ parameter.
\end{abstract}


\maketitle

\section{Introduction}
\label{sec:introduction}
The statistical properties of the cosmic microwave background (CMB)
show remarkable consistency with the paradigm of early universe inflation.
But a number of troubling anomalies persist, including the large cold spot,
the quadrupole--octupole alignment, and a hemispherical amplitude asymmetry.
If these anomalies are primordial it is not yet clear whether they can be compatible
with the simplest inflationary models
which typically predict statistical independence of each multipole
(see Ref.~\cite{Schwarz:2015cma} and references therein).
In this Letter we report results for a special set of inflationary scenarios which
can accommodate the hemispherical asymmetry.

Working with the Planck 2013 temperature data, Aiola et al. demonstrated that
the asymmetry
could be approximately fit by a position-dependent power-spectrum
at the last-scattering surface of the form~\cite{Aiola:2015rqa}
\begin{equation}
	\dimlessP^{\text{obs}}(k)
	\approx
	\frac{k^3 P(k)}{2\pi^2}
	\Big(
		1 + 2 A(k) \hat{\vect{p}} \cdot \hat{\vect{n}} + \cdots
	\Big) ,
	\label{eq:Pk-modulation}
\end{equation}
where $\hat{\vect{p}}$ represents the direction of maximal asymmetry,
$\hat{\vect{n}}$ is the line-of-sight from Earth,
and $A(k)$ is an amplitude which Aiola et al. found to scale roughly like
$k^{-0.5}$. Averaged
over $\ell \sim 2\,\text{--}\,64$
it is of order $0.07$.
In this paper our primary objective is to explain how an inflationary
model can produce an asymmetry which replicates this scale dependence.

The effect is seen in multiple frequency channels
and in the older WMAP data, which makes it less likely
to be attributable to an instrumental effect or foreground.
Future improvements in observation are likely to be driven by
polarization data, which provide an independent probe of the largest-scale
modes~\cite{Zibin:2015ccn}.

\para{Inflationary explanations}
Erickcek, Carroll and Kamionkowski proposed that~\eqref{eq:Pk-modulation}
could be produced during an inflationary epoch
if the two-point function at wavenumber
$k$ is modulated by perturbations of much
larger wavelength~\cite{Erickcek:2008sm}.
This entails the presence of a bispectrum with nonempty squeezed limit,
and if the amplitude is sufficiently large it would
be the first evidence for multiple active light fields in the inflationary era.

This is an exciting possibility but there is significant concern that
a bispectrum of this type may already be ruled out by observation.
Current experiments do not measure the bispectrum on individual
configurations,
but rather weighted averages over related groups of configurations---%
and at present are most sensitive to modestly squeezed examples.
Averaged over these configurations,
Planck observations
require
the non-Gaussian component to have amplitude
$|\fNL| / 10^5 \lesssim 0.01\%$~\cite{Ade:2013ydc,*Ade:2015ava}
compared to the leading Gaussian part.
Meanwhile, ignoring all scale dependence,
Refs.~\cite{Kanno:2013ohv,Lyth:2014mga,Namjoo:2013fka,Kobayashi:2015qma}
showed that an inflationary origin would require
\begin{equation}
	\frac{|a_{20}|}{6.9\times 10^{-6}} \frac{|\fNL|}{10} \simeq 6	
	\left(
		\frac{A}{0.07}
	\right)^2
	\beta
	\label{eq:a20-A-fNL-basic}
\end{equation}
where $a_{20}$ is the quadrupole of the CMB temperature anisotropy,
measured to be approximately
$|a_{20}| \approx 6.9 \times 10^{-6}$~\cite{Efstathiou:2003tv},
and $\beta$ is a model-dependent number
which would typically be rather larger than unity.
Therefore Eq.~\eqref{eq:a20-A-fNL-basic}
suggests that an inflationary scenario may require
$|\fNL| \gtrsim 60$,
in contradiction to measurement.
If so, we would have to abandon the possibility of an inflationary
origin, at least if produced by the Erickcek--Carroll--Kamionkowski
mechanism.
To evade this Eq.~\eqref{eq:a20-A-fNL-basic}
could be weakened by tuning our position
on the long-wavelength background
to reduce $\beta$,
but clearly we should not allow ourselves to entertain fine-tunings
which are less likely than the anomaly they seek to explain.

\para{Averaged constraints}
The requirement that $A(k)$
varies with scale
gives an alternative way out which has yet to be
studied in detail.
It could happen that the
bispectrum amplitude is large on long wavelengths
but runs to small values at shorter wavelengths in such a way
that the wavelength-averaged values measured by CMB experiments
remain acceptable.
Eq.~\eqref{eq:a20-A-fNL-basic}
might then apply for a small number of wavenumber configurations
but would have no simple relation to observable quantities.

In this Letter
we provide, for the first time, an analysis of the CMB temperature bispectrum
generated by a scale- and shape-dependent primordial bispectrum
which is compatible with the modulation $A(k)$.
We do this by constructing an explicit model which
can be contrived to match all current observations,
and also serves as a useful example
showing the complications which are encountered.
Despite its contrivance, we expect the bispectrum produced by
this model to be a good proxy for the bispectrum generated in a much
larger class of successful scenarios producing scale-dependence
through a large, negative $\eta$-parameter.
If the model can be embedded within a viable early universe scenario,
we show that it can explain the asymmetry
without introducing
tension with the $\fNL$ or low-$\ell$ amplitude constraints
(the `Grischuk--Zel'dovich' effect).

In this Letter we focus on the simplest possibility that the
non-Gaussian fluctuations of a single field
generate
the asymmetry,
although we allow a second field to generate
the Gaussian part of the curvature perturbation.
Generalizations and further details are presented
in a longer companion paper~\cite{Byrnes:2015dub}.

\section{Generating the asymmetry}
We denote the field with scale-dependent fluctuations
by $\isomode$, and take it to substantially dominate the
bispectrum for the observable curvature perturbation $\zeta$.
The $\zeta$ two-point function
$\langle \zeta(\vect{k}_1) \zeta(\vect{k}_2) \rangle
= (2\pi)^3 \delta(\vect{k}_1 + \vect{k}_2) P(k)$
can depend on $\isomode$,
or alternatively on any combination of $\isomode$ and other Gaussian fields.
The question to be resolved is how $P(k)$
responds to a long-wavelength background of $\isomode$ modes
which we write $\delta\isomode(\vect{x})$.

\para{Response function}
In Ref.~\cite{Byrnes:2015dub} we show that this
response can be computed
using the operator product expansion (`OPE'),
and expressed in terms of the ensemble-averaged
two- and three-point functions of the inflationary model.
We focus on models in which the primary effect
is due to the amplitude of the long-wavelength background
rather than its gradients. Since the perturbation is
small it is possible to write
\begin{equation}
	P(k,\vect{x}) = P(k)\Big(
		1
		+ \delta\isomode(\vect{x}) \rho_\isomode(k)
		+ \cdots
	\Big) .
	\label{eq:twopf-response}
\end{equation}
We call $\rho_\isomode(k)$ the `response function'.
It can be regarded as the derivative $\d\ln P(k) / \d \sigma$.
The OPE gives~\cite{Byrnes:2015dub}
\begin{equation}
	\rho_\isomode(k)
	\simeq
	\frac{1}{P(k)}
	[\Sigma^{-1}(k_L)]_{\isomode\lambda}
	B^\lambda(k, k, k_L)
	\quad
	\text{if $k \gg k_L$}
	,
	\label{eq:response-function}
\end{equation}
where a sum over $\lambda$ is implied,
and
$\Sigma^{\alpha\beta}$ and $B^\alpha$
are spectral functions for certain mixed two- and
three-point correlators of $\zeta$ with the light
fields of the inflationary model (and their momenta),
which we collectively denote $\delta\phi^\alpha$,%
	\footnote{In the restricted setup we are describing, where only
	$\isomode$ has a non-negligible bispectrum, the sum over
	$\lambda$ in Eq.~\eqref{eq:response-function} would include the
	field $\isomode$ and its momentum.}
\begin{subequations}
\begin{align}
	\langle \delta\phi^\alpha(\vect{k}_1) \delta\phi^\beta(\vect{k}_2) \rangle
	& = (2\pi)^3 \delta(\vect{k}_1 + \vect{k}_2) \Sigma^{\alpha\beta}
	\\
	\langle \delta\phi^\alpha(\vect{k}_1) \zeta(\vect{k}_2)
	\zeta(\vect{k}_3) \rangle
	& = (2\pi)^4 \delta(\vect{k}_1 + \vect{k}_2 + \vect{k}_3) B^\alpha .
\end{align}
\end{subequations}
Eq.~\eqref{eq:response-function}
is one of our principal new results. It enables us
to extend the analysis of inflationary models
beyond those already considered in the literature
to cases with nontrivial, scale-dependent correlation
functions. The conditions under which it
applies are discussed in more detail in Ref.~\cite{Byrnes:2015dub}.

In the special case of a 
slow-roll model in which a single field generates all perturbations, it can be shown that the right-hand
side of~\eqref{eq:response-function}
is related to the reduced bispectrum,
\begin{equation}
	\rho_\isomode(k) = \frac{12}{5} \fNL(k,k,k_L), \quad
	\text{$k \gg k_L$} ,
\end{equation}
where $\fNL(k_1, k_2, k_3)$ is defined by
\begin{equation}
	\frac{6}{5} \fNL(k_1, k_2, k_3)
	\equiv
	\frac{B(k_1, k_2, k_3)}{P(k_1) P(k_2) + \text{2 cyclic perms}} .
	\label{eq:single-field-response}
\end{equation}
Notice that,
for a generic $B(k_1, k_2, k_3)$,
the reduced bispectrum defined this way
has no simple relation
to any of the amplitudes
$\fNLlocal$, $\fNLequi$, etc., measured by experiment.
Eq.~\eqref{eq:single-field-response}
reproduces earlier results
given in the literature~\cite{Lyth:2013vha,Kanno:2013ohv,Lyth:2014mga,Kobayashi:2015qma}
but does not apply for the more realistic models
considered in this Letter.

\para{Long wavelength background}
To model the long-wavelength background we take
\begin{equation}
  \delta \isomode(\vect{x}) \approx E \dimlessP_\isomode^{1/2}(\kL)
  \cos( \vectkL \cdot \vect{x} + \vartheta )
  \label{eq:modulating-mode}
\end{equation}
where $E$ labels the `exceptionality' of the amplitude, with $E=1$
being typical and $E \gg 1$ being substantially larger than typical.
We take the wavenumber $\vectkL$ to be fixed.
The phase $\vartheta$ will vary between realizations,
and the Earth is located at $\vect{x}=0$.

The last-scattering surface is at comoving radius $\xls \approx 14,000 h \, \Mpc^{-1}$.
Evaluating~\eqref{eq:twopf-response}
and~\eqref{eq:modulating-mode} on this surface
at physical location $\vect{x} = \xls \hat{\vect{n}}$,
and assuming
$\alpha \equiv \xls \kL / 2\pi < 1$ so that
the wavelength associated with $\kL$ is somewhat larger than
$\xls$, we obtain 
\begin{equation}
	P(k,\vect{x}) = P(k)\Big(
		1
		- C(k)
		+ 2A(k) \frac{\vect{x} \cdot \hatvectkL}{\xls}
		+ \cdots
	\Big) .
	\label{eq:twopf-modulation}
\end{equation}
The quantities $A(k)$ and $C(k)$ are determined
in terms of the response $\rho_\isomode$
and long-wavelength background by
\begin{subequations}
\begin{align}
	\label{eq:A-def}
	A(k) & = \pi \alpha E \dimlessP_\isomode^{1/2}(\kL) \rho_\isomode(k) \sin \vartheta \\
	\label{eq:C-def}
	C(k) & = - A(k) \frac{\cos \vartheta}{\pi \alpha \sin \vartheta} . 
\end{align}	
\end{subequations}
Both $A$ and $C$ share the same scale-dependence, so it is possible
that $C(k)$ could be used to explain the lack of power on large scales~\cite{Lyth:2013vha,Lyth:2014mga}.
If so, the model could simultaneously explain \emph{two} anomalies---%
although this would entail a stringent constraint on $\alpha$ in order that $C(k)$
does not depress the power spectrum too strongly at small $\ell$.
The relative amplitude of $A(k)$ and $C(k)$ depends on the unknown phase
$\vartheta$ and our assumption of the form~\eqref{eq:modulating-mode},
but the observation that they scale the same way with $k$
constitutes a new and firm prediction for all models which explain the power asymmetry
by modulation from a single super-horizon mode.

\section{Building a successful model}

\para{Single-source scenarios}
In the case where one field dominates the two- and three-point
functions of $\zeta$,
the bispectrum is equal in squeezed
and equilateral configurations~\cite{Dias:2013rla,Kenton:2015lxa}.
Therefore
\begin{equation}
	\rho_\isomode=\frac{12}{5}\fNL(k,k,k_L)=\frac{12}{5}\fNL(k,k,k) ,
\end{equation}
and the asymmetry scales in the same way as the equilateral configuration
$\fNL(k,k,k)$.
If the scaling is not too large it can be computed using~\cite{Byrnes:2010ft}
\begin{equation}
	\frac{d \ln |\fNL|}{d\ln k} = \frac{5}{6\fNL} \sqrt{\frac{r}{8}}\frac{\Mp^3 V''' }{3H^2} ,
\label{eq:nfnl-ss}\end{equation}
where $r \lesssim 0.1$ is the tensor-to-scalar ratio.
To achieve strong scaling we require $\Mp^3 V'''/(3H^2)\gg1$.
But within a few e-foldings this
will typically generate an unacceptably large
second slow-roll parameter $\eta_\isomode$, defined by
\begin{equation}
    \eta_\isomode= \frac{\Mp^2 V''}{3H^2} .
    \label{eq:eta-isomode}
\end{equation}
Therefore it will
spoil the observed near scale-invariance of the power spectrum. 

As a specific example, a self-interacting curvaton model
was studied in Ref.~\cite{Byrnes:2015asa}.
This gave rise to many difficulties,
including logarithmic running of $\fNL(k,k,k)$ with $k$---%
which is not an acceptable fit to the scale dependence of $A(k)$~\cite{Aiola:2015rqa}.
Even worse, because~\eqref{eq:nfnl-ss}
is large only when $\fNL$ is suppressed below its natural value,
both the trispectrum amplitude $\gNL$ and the
quadrupolar modulation of the power spectrum
were unacceptable.
In view of these difficulties we will not pursue single-source models further.

\para{Multiple-source scenarios}
In multiple-source scenarios there is more flexibility.
If different fields contribute to the power spectrum and bispectrum
it need not happen that a large $\eta_\isomode$
necessarily spoils scale-invariance.
In these scenarios
$\rho_\isomode$ no longer scales
like the reduced bispectrum, but rather its square-root
$\fNL(k,k,k)^{1/2}$.
Therefore
\begin{equation}
\begin{split}
	\frac{\d\ln A}{\d\ln k}
	& \approx
	\frac{1}{2} \frac{\d\ln |\fNL(k,k,k)|}{\d\ln k}
	\\
	& \approx
	\frac{\d\ln(\dimlessP_\isomode / \dimlessP) }{\d\ln k}
	\approx 2 \eta_\isomode - (n_s-1) 
	\label{eq:scalings}
\end{split}
\end{equation}
where $\dimlessP$ is the dimensionless power spectrum,
$n_s-1\simeq-0.03$ is the observed scalar spectral index
and $\eta_\isomode$ was defined in Eq.~\eqref{eq:eta-isomode}.
If we can achieve a constant $\eta_\isomode \approx -0.25$ while
observable scales are leaving the horizon then it is possible
to produce an acceptable power-law for $A(k)$.
For further details of these scaling estimates for $A(k)$
see Kenton et al.~\cite{Kenton:2015jga} or Ref.~\cite{Byrnes:2015dub}.

A simple potential with large constant $\eta_\isomode$ is
\begin{equation}
	W(\phi,\isomode)
	=
	V(\phi)
	\left(
		1-\frac{1}{2} \frac{m_\isomode^2 \isomode^2}{\Mp^4}
	\right) .
\end{equation}
The inflaton $\phi$ is taken to dominate the energy density and therefore
drives the inflationary phase.
Initially $\isomode$ lies near the hilltop at $\isomode=0$,
so its kinetic energy is subdominant and
$\epsilon \approx \Mp^2 V_\phi^2 / V^2$. (Here
$\epsilon = -\dot{H}/H^2$ is the conventional slow-roll parameter.)
As inflation proceeds $\isomode$ will roll down the hill
like $\isomode(N) = \isomode_\star \e{-\eta_\isomode N}$,
where `$\star$' denotes evaluation at the initial time and
$N$ measures the number of subsequent e-folds.

To keep the $\isomode$ energy density subdominant we must
prevent it rolling to large field values,
which implies that $\isomode_\star$ must be chosen
to be very close to the hilltop.
But the initial condition must also lie outside the diffusion-dominated
regime, meaning the classical rolling should be substantially larger
than quantum fluctuations in $\isomode$.
This requires
$|\d\isomode/\d N| \gg H_\star / 2\pi$.
In combination with the
requirement that $\isomode$ remain subdominant in the observed power spectrum,
we find that $\isomode_\star$ should be chosen so that
$|\isomode_\star| \gtrsim  \sqrt{\epsilon_\star \dimlessP}\Mp / |\eta_\isomode |$.
For typical values of
$\epsilon = 10^{-2}$ and $\eta_\isomode =-0.25$
this requires $|\isomode(60)| \gtrsim 100 \Mp$
which is much too large.
The problem can be ameliorated by reducing $\epsilon_\star$,
but then $\isomode$ contributes significantly to
$\epsilon$ during the inflationary period.
This reduces the bispectrum amplitude to a tiny value,
or causes $\isomode$ to contaminate the power spectrum and
spoil its scale invariance~\cite{Byrnes:2015dub}.

\para{A working model}
To avoid these problems, consider a potential in which the
effective mass of the $\isomode$ field makes a
rapid transition.
An example is
\begin{equation}
	W=W_0\left(1+\frac12\eta_\phi\frac{\phi^2}{\Mp^2}\right)\left(1+\frac12\eta_\isomode(N)\frac{\isomode^2}{\Mp^2}\right) ,
	\label{eq:tanh-model}
\end{equation}
where $\eta_\isomode(N)$ is chosen
to be $-0.25$ while observable scales exit the horizon,
later running rapidly to settle near $-0.08$.
(For a concrete realization see Ref.~\cite{Byrnes:2015dub}.)
We take the transition to occur
roughly 16 e-folds after the largest observable
scales exited the horizon.
The field $\phi$ will dominate the Gaussian part of $\zeta$
and its mass  should be chosen to match the observed spectral index.

Although simple and illustrative, this model is not trivial to embed in
a fully realistic early universe scenario.
The required initial value of $\isomode$ is only
a little outside the quantum diffusion regime
which may lead to unwanted observable consequences.
Also, not all isocurvature modes decay by the end of the inflationary epoch
so~\eqref{eq:tanh-model} should be completed by a specification
of the reheating model, and it is possible this could change the
prediction for the $n$-point functions.
But, assuming these problems are not insurmountable,
we can accurately compute the bispectrum generated by~\eqref{eq:tanh-model}.
Our predictions then apply to any successful realization of this
scenario.

\para{Estimator for $\fNLlocal$}
The most urgent question is whether the bispectrum
amplitude is compatible with present constraints
for $\fNLlocal$, $\fNLequi$, etc., which as explained
above are weighted averages over the bispectrum amplitude on
groups of related configurations.
At present the strongest constraints apply to $\fNLlocal$
which averages over modestly squeezed configurations.

To determine the response of these estimators
we construct a Fisher estimate.
We numerically compute $\sim 5 \times 10^6$
bispectrum configurations for~\eqref{eq:tanh-model}
covering the range from $\ell \sim 1$ to $\ell \sim 7000$
and use these to predict the
observed angular temperature bispectrum.

For a choice of parameter values which generate the correct
amplitude and scaling of $A(k)$, we find that a Planck-like
experiment would measure
order-unity values,
\begin{equation}
  \hatfNLlocal = 0.25
  ,
  \quad
  \hatfNLequi = 0.6
  ,
  \quad
  \hatfNLortho = -1.0
  .  
  \label{eq:fNL-estimator-predictions}
\end{equation}
These estimates are our second principal result.
They are one to two orders of magnitude smaller than
previous estimates based on Eq.~\eqref{eq:a20-A-fNL-basic},
and are easily compatible with present-day constraints.
The difference comes from the strong running
of the bispectrum amplitude required for compatibility with
$A(k)$,
and also the growing number of bispectrum configurations
available at large $\ell$.
This means that the signal-to-noise tends to be dominated
by the largest-$\ell$ configurations where the amplitude
is small, depressing the final weighted average;
in fact,
we find that the reduced bispectrum amplitude near the
Planck pivot scale $\ell \sim 700$ is a fair predictor
for the averages~\eqref{eq:fNL-estimator-predictions}.

We find that
it is possible to simultaneously satisfy observational constraints on
the amplitude of low-$\ell$ multipoles of the power spectrum~\cite{Byrnes:2015dub}.
For example, choosing
$\alpha=0.01$ (which makes the wavelength of the modulating mode roughly 100 times the
distance to the last-scattering surface)
requires an exceptionality $E \approx 300$
to match the measured amplitude of $A(k)$.
A value for $E$ in this range would likely require \emph{further} new physics,
but it could perhaps be reduced to a value of order $10$ by increasing the bispectrum amplitude.
For these parameter choices the low-$\ell$ suppression $C(k)$
may be larger than the
approximate bound $C(k) \lesssim 0.14$ suggested by Contaldi et al.~\cite{Contaldi:2014zua}.
(There is some uncertainty regarding the precise numerical bound, because
the result of Contaldi et al. assumed the BICEP measurement of $r$ which is now known
to have been confused by dust.)
If necessary this would apparently have to be mitigated
by tuning our position on the long-wavelength mode.

Finally, we note that
although the precise bispectrum used in our analysis
applies to the specific model~\eqref{eq:tanh-model}, any
model which generates scale dependence through a large $\eta_\isomode$
is expected to produce a similar shape.
Therefore, despite the contrivances of our example,
we expect our conclusions to be robust and apply much more generally.

\section{Conclusions}
The CMB power asymmetry is a puzzling feature which may impact on
our understanding of the very early universe.
The most popular inflation-based explanations deploy
the Erickcek--Carroll--Kamionkowski mechanism,
in which a single super-horizon mode of exceptional amplitude
modulates the small-scale power spectrum (see Ref.~\cite{Adhikari:2015yya} for a generalization to include all superhorizon modes).
But until now, comparisons of the scenario with observation
have not accounted for the scale-dependence of the asymmetry---%
or the bispectrum which is responsible for it.
This is a necessary feature of the model.
Previous analyses based on Eq.~\eqref{eq:a20-A-fNL-basic} have suggested
the required bispectrum amplitude may be incompatible with
observation, but it has not been clear how the inclusion of
scale-dependence would modify this conclusion.

In this Letter we have presented a direct determination of
the response function which couples the asymmetry to
the ensemble-averaged bispectrum and the super-horizon mode.
We have presented an illustrative example which satisfies
all current observational constraints,
and which can be used to obtain precise predictions for the primordial
bispectrum.
Using this to predict the angular bispectrum of the CMB temperature
anisotropy we have confirmed that the bispectrum amplitude
is well within the bounds set by current Planck data.

Although this bispectrum strictly applies for the step model~\eqref{eq:tanh-model}
we believe it to be a good proxy for any inflationary explanation of the asymmetry
which uses a large $\eta$ parameter to generate the scale dependence.
Our results show that such scenarios involve much less tension with
observation than would be expected on the basis of~\eqref{eq:a20-A-fNL-basic}.
Nevertheless, this does not mean that an inflationary explanation is
automatically attractive.
To build a successful model we have been forced to make a number of arbitrary
choices, including the initial and final values of the $\isomode$ mass,
and the location and rapidity of the transition.
It is also unclear whether this inflationary model can be embedded
within a viable early universe scenario,
which should include at least
initial conditions for the inflationary era and a description of
how reheating connects it to a subsequent radiation epoch.
In our present state of knowledge it seems challenging to construct
a scenario including all these
features, and capable of explaining the hemispherical asymmetry,
which does not involve choices at least as unlikely
as the asymmetry itself.

\subsection*{Acknowledgements}
DS acknowledges support from the Science and Technology
Facilities Council [grant number ST/L000652/1].
CTB is a Royal Society University Research Fellow.
The research leading to these results has received funding from
the European Research Council under the European Union's
Seventh Framework Programme (FP/2007--2013) / ERC Grant
Agreement No. [308082]. This work was supported in part by
National Science Foundation Grant No. PHYS-1066293.

\para{Data availability statement}
Please contact the authors
to obtain the bispectrum for the step model~\eqref{eq:tanh-model},
which was used to estimate the responses~\eqref{eq:fNL-estimator-predictions}.

\bibliography{refs}
\end{document}